\documentstyle[preprint,epsf,aps]{revtex}
\begin{document}
\def\q {\tilde{q}}
\def\t {\widetilde{t^{c}}}
\def\tt {\tilde{t}}
\def\H {\widetilde{H}}
\def\bt {\bar{t}}
\def\blam {\bar{\lambda}}
\def\cM {{\cal M}}
\def\beq {\begin{equation}}
\def\eeq {\end{equation}}
\def\half {{1\over 2}}
\def\third {{1\over 3}}
\draft
\preprint{\vbox{\hbox{KUNS-1351} \hbox{HE(TH) 95/11}}}
\title{Parameter Bounds in the Supersymmetric Standard Model
from Charge/Color Breaking Vacua}
\author{Andrew J. Bordner
\thanks{e-mail address: bordner@gauge.scphys.kyoto-u.ac.jp}}
\address{Department of Physics, Kyoto University, Kyoto 606-01, Japan}
\date{\today}
\maketitle
\begin{abstract}
We calculate limits on the trilinear soft-breaking
    parameter, $A_{t}$, in the Minimal
    Supersymmetric Standard model by requiring the absence of nonzero
    top squark vacuum expectation values.
Assuming a low $\tan\beta$,
    which implies a large top Yukawa coupling, we also calculate one-loop
    corrections to the effective potential.  The resulting numerical
    calculations of the charge/color breaking limits are presented as
    best-fit surfaces. We compare these results with the analytical
    limit, $A_{t}^{2} < 3(m_{2}^{2}+m_{\q}^{2}+m_{\tt}^{2})$,
    and find that although this is a good estimate of the charge/color
    breaking bounds for a simplified model of the top sector,
    stricter bounds are found by a numerical minimization of the
    Minimal Supersymmetric Standard Model potential.
\end{abstract}
\section{Introduction}
The Minimal Supersymmetric Standard Model (MSSM) may be described by a
Lagrangian
containing interactions consistent with invariance under the gauge
group $SU(3) \times SU(2) \times U(1)_{Y}$ and global supersymmetry plus
a Lagrangian containing a restricted set of soft supersymmetry
breaking terms \cite{MSSM}.  These terms break supersymmetry
while
maintaining a useful property of a supersymmetric theory, namely the
cancellation of quadratic divergences \cite{soft breaking}.  The
absence of
these
divergences is necessary in order to define the renormalized mass of a
fundamental scalar, such as the Higgs boson, without a fine-tuning of
the cancellation between the bare mass and the scalar
self-energy \cite{Nilles}.

The presence of fundamental scalar fields in the MSSM, besides the
Higgs
bosons, leads to the possibility that these fields may acquire
non-zero
vacuum expectation values (vevs).  Since this would violate the
conservation of color and/or electric charge symmetry, this leads to
forbidden regions of the parameter space of the theory.  We will
calculate numerical estimates of the boundary of the allowed region of
soft-breaking parameters using both the tree-level potential and the
one-loop effective potential.

Many studies of the MSSM mass spectrum neglect these charge/color
breaking, or CCB, bounds in their analyses.  Previously, CCB bounds were
obtained for various supersymmetric models, however no systematic
numerical study of CCB constraints for
a realistic approximation to the MSSM using the one-loop effective
potential has been done \cite{previous CCB,GHS_model}.

One may assume that there are relations among
the soft breaking terms, such as in the minimal supergravity model in which
all scalar masses and scalar trilinear couplings are the same at the
unification scale, of order $10^{16}~GeV$ \cite{Low-E SUGRA}.  However we will
find constraints on the
soft-breaking parameters at a low-energy scale, $Q_{0}$, with
$Q_{0} < 1~TeV$.  This is an
indeterminate upper limit on particle masses if the MSSM is to explain
the gauge hierarchy problem.  We will not make any assumptions about
the theory near the GUT scale nor the particle spectrum
above $Q_{0}$.

We will use an approximation to the MSSM that includes only the top
flavor supermultiplets.  This follows from evidence that the top quark
mass $m_{t} \approx 176~GeV$ \cite{top quark mass}.  We use the conventional
definition
$\tan\beta\equiv v_{2}/v_{1}$, with $v_{1}$, $v_{2}$ the vevs
for the Higgs scalar fields, $H_{1}$ and $H_{2}$, respectively.
Assuming
a small value for $\tan\beta$, near $1.0$, gives
the top quark Yukawa
coupling, $h_{t} = 1.0$.
The contributions from the bottom supermultiplets may then be ignored.

There are various reasons to choose these particular values of
$\tan\beta$ and to consider only the top squarks as acquiring a
non-zero vev.  First of all, there is an infrared quasi-fixed
point in the renormalization group equation for $h_{t}(Q)$ which
corresponds to a value
$h_{t}^{FP}(m_{t})\approx 1.25$ \cite{IR_fixed_pt}.  The
mass relation
\beq
m_{t}(m_{t}) = {h_{t}^{FP}(m_{t})\over \sqrt(2)}v\sin\beta
\eeq
gives $\tan\beta\approx 1.2$ if one uses the relation between the
top quark mass defined by the pole in its propagator and its
running mass to first-order in
the QCD gauge coupling \cite{pole mass corr.}.
\beq
m_{pole} = m_{t}(m_{t})\left[1+{4\over3\pi}\alpha_{3}(m_{t})
+{\cal O}(\alpha_{3}^{2})\right].
\eeq
Therefore a value of $\tan\beta$
at $Q_{0}$ in the range $1.1 < \tan\beta < 2.0$ results from
a large range of $h_{t}$ values at the GUT scale.  Although
$\tan\beta$ is not required to be in this range, it
indicates that this is a natural choice.  One motivation
for considering only the top sector comes from assuming common
soft-breaking parameters at the GUT scale.  A large value of $h_{t}$
causes the third generation parameters to undergo the largest change
as they are evolved from $Q_{GUT}$ down to $Q_{0}$.
For this same reason, $h_{t}$ also gives the largest contribution to the
radiative gauge symmetry breaking of $SU(2) \times U(1)_{Y}$
\cite{radiative breaking,GRZ}.  Therefore, if one assumes that the minimum of
the effective potential at energy scales $Q \gg Q_{0}$
gives zero vevs for the scalar fields, such as in the case of
universality at the GUT scale, as one evolves to $Q_{0}$ the
third-generation parameters undergo the largest change
and the CCB constraints from third generation scalar
fields will be the most restrictive.
Finally, as discussed in Ref. \cite{tunneling},
the potential barrier height for
tunneling from the the symmetric vacuum at a high temperature
($T > 100~GeV$), early in the expansion of the universe, to
a lower CCB minimum is proportional to $1/h_{min}^{2}$ where
$h_{min}$ is the smallest of the Yukawa couplings for the slepton and
squark fields that have non-zero vevs at the CCB minimum.  This
implies that one should consider CCB vacua in which only the Higgs
fields and the top squarks have non-zero vevs in order for the tunneling
from the symmetric to the CCB vacuum to have occurred in a time
less than the present
age of the universe ($\approx 10^{10}$ years).
\section{Approximation to the MSSM}
We use a consistent approximation to the MSSM with $\tan\beta=1.5$
as a small value near the fixed point value and interactions with the
bottom quark superfields are ignored.  We use all MSSM interactions
between the following fields $H_{1}$, $H_{2}$, $\H_{1}$,
$\H_{2}$, $q$, $t^{c}$, $\q$, $\t$, $A_{\mu}$, and $\lambda$.
$H_{1}$ and $H_{2}$ are respectively the hypercharge $-1$ and $+1$
Higgs boson doublets.  The corresponding field variables with a tilde are
the Higgsino doublets.  $q$ and $t^{c}$ are the left-handed component
of the top quark and the right-handed component of the charge
conjugate top quark field respectively.  Again, the corresponding field
variables with tildes are the top squarks.  $A_{\mu}$ is the gluon
field and $\lambda$ is the gluino field.  Notice that the field
content in this approximation is supersymmetric.  This arises from
including all interactions with the top quark supermultiplet involving
the parameters $h_{t}$ and $\mu$.  The potential in this approximation
as well as the definitions of the parameters appearing in it are shown
in the appendix.

We use the values of the gauge couplings at the weak
scale $M_{Z}$: $g_{1}=0.358$, $g_{2}=0.652$, and
$g_{3}=1.213$
for all calculations \cite{PDG}.  Since we will take the
renormalization scale to be $Q_{0} = 500~GeV$, the running of
the gauge couplings from $M_{Z}$ to this scale is
negligible.
We also omit the couplings $g_{1}$ and $g_{2}$ in the calculation of
the one-loop effective potential.  However, we retain the terms with
these couplings at tree-level since they include the quartic Higgs
scalar interactions responsible for the non-zero Higgs vev of the
Standard Model (SM).
\section{The Effective Potential}
Quantum corrections may affect whether spontaneous symmetry breaking
occurs in a field theory.  The vevs of the scalar fields are the
values of the classical fields at the minimum of the effective
potential.  We use the one-loop correction to the effective potential
in the dimensional reduction, $\overline{DR}$, renormalization
scheme \cite{DR scheme}
\beq
\label{one-loop Veff}
V_{eff}^{1-loop} = {1\over {64\pi^{2}}}Str
\left[\cM^{4}\left(\ln {\cM^{2}\over Q^{2}} - {3\over 2}\right)
\right],
\eeq
where the supertrace, $Str$, is over all color and spin degrees of
freedom with a minus sign for fermions.  $Q^{2}$ is the square
of the renormalization scale, which we take to be equal
to $Q_{0}^{2}$. $\cM^{2}$ is the field dependent mass matrix; for
instance, for a theory with scalar fields ${\phi_{i}, i = 1,\ldots, n}$
\beq
\cM_{ij} =
\left.{\partial^{2}V\over{\partial\phi_{i}\partial\phi_{j}}}
\right|_{\phi=\phi_{c}}.
\eeq
$\cM$ as well as $V_{eff}$ are function of the classical fields
$(\phi_{i})_{c} = \left\langle 0|\phi_{i} |0\right\rangle$.
Since the top Yukawa coupling, $h_{t}$, and the strong gauge coupling,
$g_{3}$, are large one may expect significant contributions to the
effective potential from one-loop corrections.  The one-loop
effective potential
is also more stable under a change of renormalization
scale \cite{GRZ}.  Since the parameters appearing in the equation for the
one-loop correction to $V_{eff}$ are the renormalized parameters we
consider the resulting CCB constraints as being limits on these
renormalized parameters in the $\overline{MS}$ scheme at momentum
scale $Q_{0}$.  If large logs of ratios of massive parameters
appear in the effective potential, such as
$\ln\left(m_{i}^{2}/ m_{j}^{2}\right)$ or
$\ln\left(\phi_{i}^{2}/ m_{j}^{2}\right)$, then one may, in
  principle, sum these using the renormalization group
equation \cite{RGE improved Veff}.  However, this is difficult when there are
multiple scalar fields and masses.  We only consider
masses, $m_{i}$, and the
renormalization scale, $Q_{0}$, which differ by less than two orders
of magnitude.  Furthermore, since we are only interested in the
effective potential near its minimum where $(\phi_{i})_{c} \sim m_{j}$,
i.e., the classical field near the minimum of $V_{eff}$ is of the same
order of magnitude as the masses, the logarithms appearing in
$V_{eff}$ are small.  Renormalization group improvement of the
effective potential in this case is unnecessary.

We also assume that the Lagrangian parameters are real.
In a field basis where the Higgs fields
are real and positive and $h_{t}$ and the scalar masses, $m_{i}^{2}$,
real, only one of the parameters
$A_{t}$, $\mu$, or $m_{\lambda}$ may be made real by redefining the
fields.
However, it was shown in Ref. \cite{Real Pars} that the presence of
complex phases in these parameters, greater than about $10^{-2}$,
gives too large of a contribution to the neutron's electric dipole
moment due to gluino loops.
\section{Numerical Calculation}
We evaluate the one-loop effective potential using Eq.
(\ref{one-loop Veff}) by diagonalizing the mass matrix $\cM$.
If one chooses the Landau gauge for the gluons, in which the propagator
is
\beq
-\imath\left[g_{\mu\nu} - {k_{\mu}k_{\nu}}\over{k^{2}}\right]
{1\over{k^{2}+\imath\epsilon}},
\eeq
the generalized mass
matrix $\cM$ is block diagonal in the gluon fields, bosonic fields,
and fermionic fields.   The top squark fields may also be rotated by
a global $SU(3)$ transformation so that they are of the form
\beq
\q_{i} = \left( \begin{array}{c} q_{1R} \\
0 \\ 0 \end{array}\right),
\qquad
\t_{i} = \left( \begin{array}{c} t_{1R} + \imath t_{1I} \\
t_{2R} \\ 0 \end{array}\right),
\eeq
with $q_{1R}$, $t_{1R}$, $t_{1I}$, and $t_{2R}$ are real.
We then use only the real part of the one-loop effective potential;
the imaginary part is related to the decay rate of the vacuum as shown
in Ref. \cite{Weinberg and Wu}.

The global minimum of the effective potential is found by calculating
the local minimum using a standard algorithm from
Ref. \cite{Numerical Recipes} starting from a set of field values on a
rectangular grid.  If this global minimum occurs for non-zero values
of the squark fields then there is a CCB vacuum.  This process is
repeated for a set of soft-breaking parameter values on a rectangular
grid in parameter space.  A quadratic surface is then fit to
the parameter points on the boundary between those that give a
symmetric vacuum and those that lead to a CCB vacuum.  The surface
is fit to the boundary points by varying the coefficients in the
equation of the surface, $(P_{1},\ldots,P_{8})$, until the
average of the
distance squared from the boundary points to the nearest point on the
surface, $\langle d^{2}\rangle$, is minimized.  The equation of the
surface is
\beq
\label{quadratic_surface}
A_{t}^{4}+P_{1}A_{t}^{2}+P_{2}m_{2}^{4}+P_{3}m_{2}^{2}
+P_{4}m_{\q}^{4}+P_{5}m_{\q}^{2}+P_{6}m_{\tt}^{4}+P_{7}m_{\tt}^{2}
-P_{8}=0
\eeq
with $(A_{t},m_{2}^{2},m_{\q}^{2},m_{\tt}^{2})$ a
point in parameter
space.  These soft-breaking
parameters are defined in the appendix.  This particular
parameterization of the surface is chosen solely because of its
simplicity, but as will be seen later it gives an accurate characterization of
the CCB boundary.

Because of the large dimension of the parameter space in which we want
to find CCB bounds, it is necessary to initially constrain some of the
parameters values in order to reduce the calculation time.
Two of the soft-breaking parameters, $m_{1}^{2}$ and $m_{3}^{2}$, are
constrained using two different methods.  In method S the values of
these parameters are chosen such that there is a Standard Model
minimum in the effective potential with $\langle \q\rangle =
\langle \t\rangle = 0$ and $\langle H_{1}\rangle = v\cos\beta$,
$\langle H_{2}\rangle = v\sin\beta$ with $v=246~GeV$.  This is done by
solving the
analytical expression for the one-loop effective potential with zero
vevs for the top squarks in terms of these
parameters.  These relations are shown in Eq. (\ref{one_loop_m1m3}).  In the
other
procedure, method F, we
simply fix $m_{1}^{2}$ and
$m_{3}^{2}$ at constant values, i.e., independent of the other
soft-breaking parameters.

\section{Results}
\subsection{CCB Bounds for a Simplified Model}
\label{simplified model}
We begin with a numerical analysis of the model with three real scalar
fields $H$, $Q$, and $T$ of Ref. \cite{GHS_model}.
The potential in this model is
\begin{eqnarray}
\label{GHS_pot}
h^{2}V &=& (Q^{2}T^{2} + Q^{2}H^{2} + T^{2}H^{2}) - 2AQTH +
m_{Q}^{2}Q^{2} \\
&+& m_{T}^{2}T^{2} + m_{H}^{2}H^{2}
+ {1\over {8h^{2}}}\Biggl[ g_{1}^{2}(H^{2} + \third Q^{2}
- {4\over 3}T^{2})^{2} \nonumber \\
&+& g_{2}^{2}(H^{2} - Q^{2})^{2}
+{4\over 3}g_{3}^{2}(Q^{2} - T^{2})^{2}\Biggr] \nonumber
\end{eqnarray}
This is a simplified version of the potential described in the
appendix with $Q = \q$, $T = \t$, and $H = H_{2}^{0}$.
The CCB bound
\beq
\label{analytical_bound}
A^{2} < 3(m_{2}^{2} + m_{Q}^{2} + m_{T}^{2}),
\eeq
derived in Ref. \cite{GHS_model} follows from minimizing the potential
only in the equal-field direction, {\it i.e.}, $Q=T=H$.
This is valid, in general, only if the D-term dominates, or if
$g_{i}/h \gg 1$ with $g_{i}$ the smallest gauge coupling.

We perform a numerical analysis of the potential in Eq. (\ref{GHS_pot})
using the parameter value ranges shown in Table \ref{GHS_range} and
with $h = 0.1$.  The gauge couplings are fixed at their running values
at the weak scale $M_{Z}$ as stated earlier.  The resulting best fit
to the CCB boundary is shown in Table \ref{GHS_fits}.
Clearly, this is close
to the analytical result of Eq. (\ref{analytical_bound}), as it should
be with a small value for $h$.

Next we repeat the analysis for the potential of Eq. (\ref{GHS_pot})
with the same soft-breaking parameter ranges of
Table \ref{GHS_range}, but we set $h = 1.0$.  This is the proper
value for the top quark Yukawa
coupling for small $\tan\beta$.  The corresponding CCB bound is given in
Table \ref{GHS_fits}.

The CCB bound for $h = 1.0$ is more stringent than the bound
for $h = 0.1$, i.e., for a
given set of parameter values within the range given in
Table \ref{GHS_range}, $(m_{2}^{2},m_{\q}^{2},m_{\tt}^{2})$, the
value of the $A$ parameter at which the vacuum becomes CCB is lower.
The fractional difference in the $A$ parameter bound between these
cases is largest when the other parameters, $(m_{2}^{2},m_{\q}^{2},
m_{\tt}^{2})$, are near the lower end of their
range in Table \ref{GHS_range}, where $\Delta A/A \approx 15$.
Also the value of $A$ at the CCB bound in this case is small.
The average fractional difference over the entire parameter ranges
shown in Table \ref{GHS_range} for the two bounds is
$\Delta A/A \approx 0.14$

\subsection{CCB Bound for the MSSM}
Next we examine the CCB bounds obtained from
of our approximation to the MSSM with the Lagrangian given in the appendix.
In all of the following analyses the mass units are $TeV$. We also
fix $\tan\beta = 1.5$ and $\mu = 0.4$.  The renormalization scale for
the one-loop calculation
of the effective potential is set at $Q=0.5$.

We first examine the bound for method F with $m_{1}^{2} = 0.65$ and
$m_{3}^{2} = 0.40$.  These masses result in a value of
$m_{2}^{2}=0.26$ if a S.M. minimum were required, i.e., zero squark
vevs.  These values for $m_{1}^{2}$ and $m_{3}^{2}$ were chosen
since they are consistent with a S.M. minimum for $\tan\beta = 1.5$ and
they are close to the renormalization scale, $Q = 0.5$.
The ranges of parameter values for which the effective potential was
calculated are shown in Table \ref{MSSM_range}.
The CCB bounds for the analyses of both the tree-level and the one-loop
effective potential are shown in Table \ref{method_F_fits}.
Parameter values which give a potential that is unbounded
below are not used in
determining the best fit surface for the CCB bound.
This includes
values with $m_{2}^{2} < 0.15$ for the tree-level potential since
in this case the
potential is unbounded
for $m_{2}^{2} < 2m_{3}^{2} - m_{1}^{2}$ and zero squark vevs.  We
do not examine negative $m_{2}^{2}$ values since in this case,
for the small range
of $m_{\q}^{2}$ and $m_{\tt}^{2}$ values for which the potential is
bounded below, the vevs for the C.C.B. vacuum are too small for
the numerical methods to distinguish this minimum from the minimum
with all vevs vanishing.  However, according to
Eq. (\ref{Higgs_mass_def}), the Higgs soft-breaking mass, $m_{H2}^{2}$,
is negative over part of the parameter
range examined.

Next we calculate the CCB bounds using method S, in which $m_{1}^{2}$
and $m_{3}^{2}$ are fixed by requiring a S.M. minimum.  The
parameter values listed in Table \ref{MSSM_range} were used in
determining the bounds.  Since there is no Standard Model minimum for
$m_{2}^{2} < (60~GeV)^{2}$, we do not examine $m_{2}^{2}$ values in this
range.  The CCB bounds calculated by finding the global
minimum of the both the tree-level potential and the one-loop
effective potential are given in Table \ref{method_S_fits}.

\section{Discussion}

The quadratic surface of Eq. (\ref{quadratic_surface}) is sufficient to
provide an accurate characterization of the numerical CCB bound.  All
of the least-square fits give the average distance
squared, $\langle d^{2} \rangle \leq 1.0 \times 10^{-2}$.  The one-loop
CCB bound calculations yield a larger value for $\langle d^{2}
\rangle$ because the longer calculation time requires using less
parameter grid points than for the tree-level calculation.
Typically it takes around $20$ days of cpu time on a Digital Alpha
workstation to calculate the one-loop CCB boundary points and perform
a least-squares fit so we are limited by the computation time.

As stated in section \ref{simplified model},
the numerical CCB bound for the simplified Lagrangian of
Ref. \cite{GHS_model} with Yukawa coupling $h=1.0$ is significantly
different from the numerical bound with $h=0.1$, and hence also different
from the analytical bound of Eq. (\ref{analytical_bound}), when the
other soft-breaking parameters are small.  However, for the remainder
of the parameter values tested, the numerical bound with $h=1.0$ is
quite close to the analytical bound.
One would not
expect
the CCB bound of
Eq. (\ref{analytical_bound}) to be correct for large $h$.
One possible
explanation is that the $g_{3}^{2}$ D-term is large enough at least to
insure that the minimum is in the direction $Q=T$.
Also for
$h=1.0$ the numerical bound does give a more stringent CCB
bound than that of Eq. (\ref{analytical_bound}) over the
entire range of parameters tested.

\begin{figure}
\epsfxsize=7.0in
\epsfbox{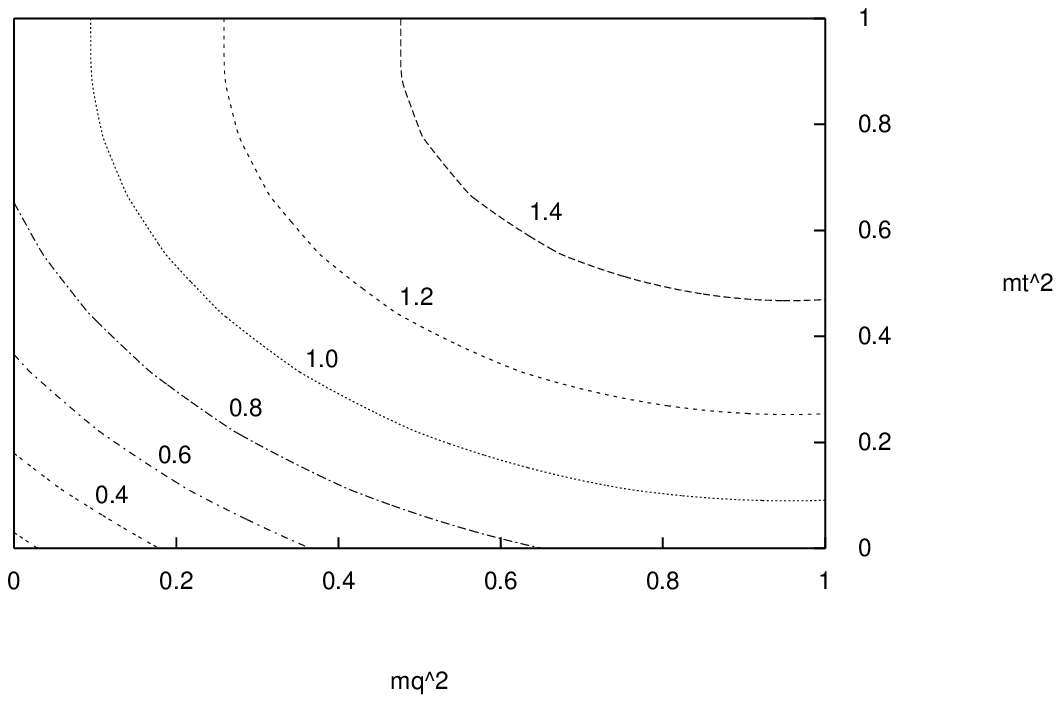}
\caption{Tree-level CCB bound for method F given in
Table \protect\ref{method_F_fits}.  $m_{2}^{2}=0.25$ and
each contour is labeled with the corresponding maximum $A$ value.}
\label{tree_F_fig}
\end{figure}

\begin{figure}
\epsfxsize=7.0in
\epsfbox{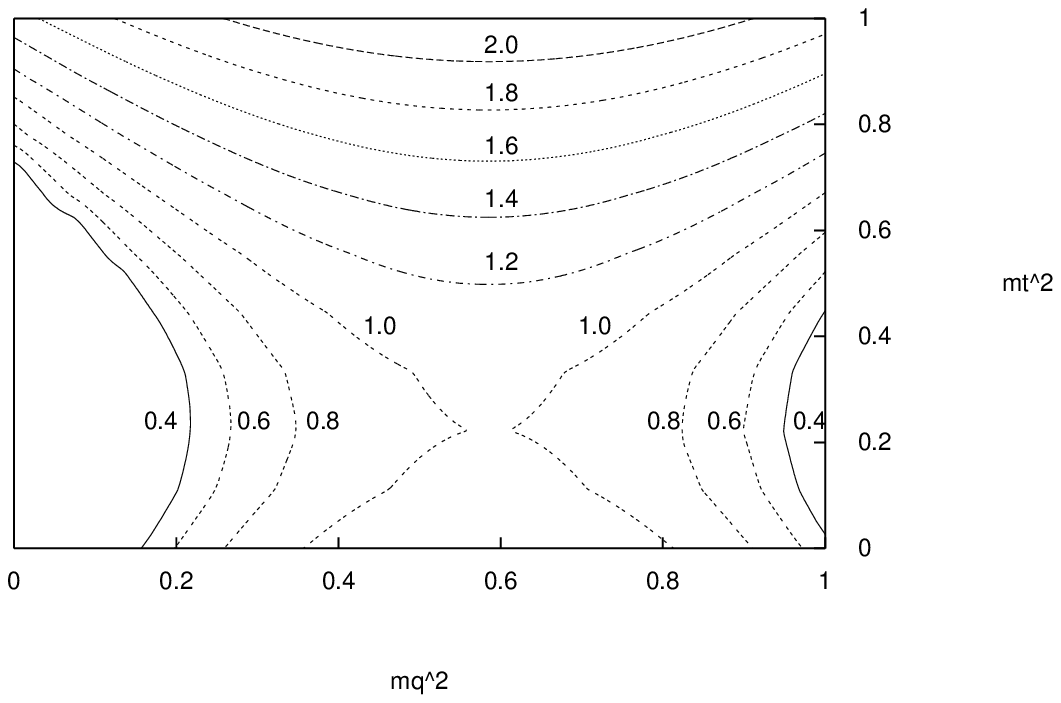}
\caption{One-loop CCB bound for method F given in
Table \protect\ref{method_F_fits}.  $m_{2}^{2}=0.25$ and
each contour is labeled with the corresponding maximum $A$ value.}
\label{loop_F_fig}
\end{figure}

\begin{figure}
\epsfxsize=7.0in
\epsfbox{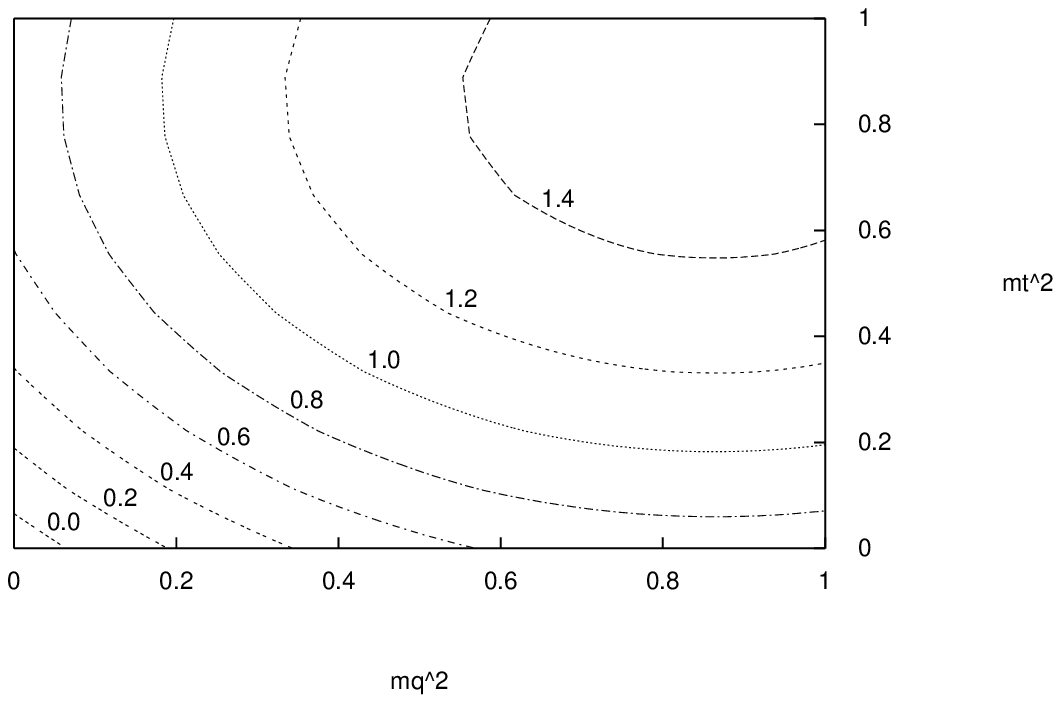}
\caption{Tree-level CCB bound for method S given in
Table \protect\ref{method_S_fits}.  $m_{2}^{2}=0.25$ and
each contour is labeled with the corresponding maximum $A$ value.}
\label{tree_S_fig}
\end{figure}

\begin{figure}
\epsfxsize=7.0in
\epsfbox{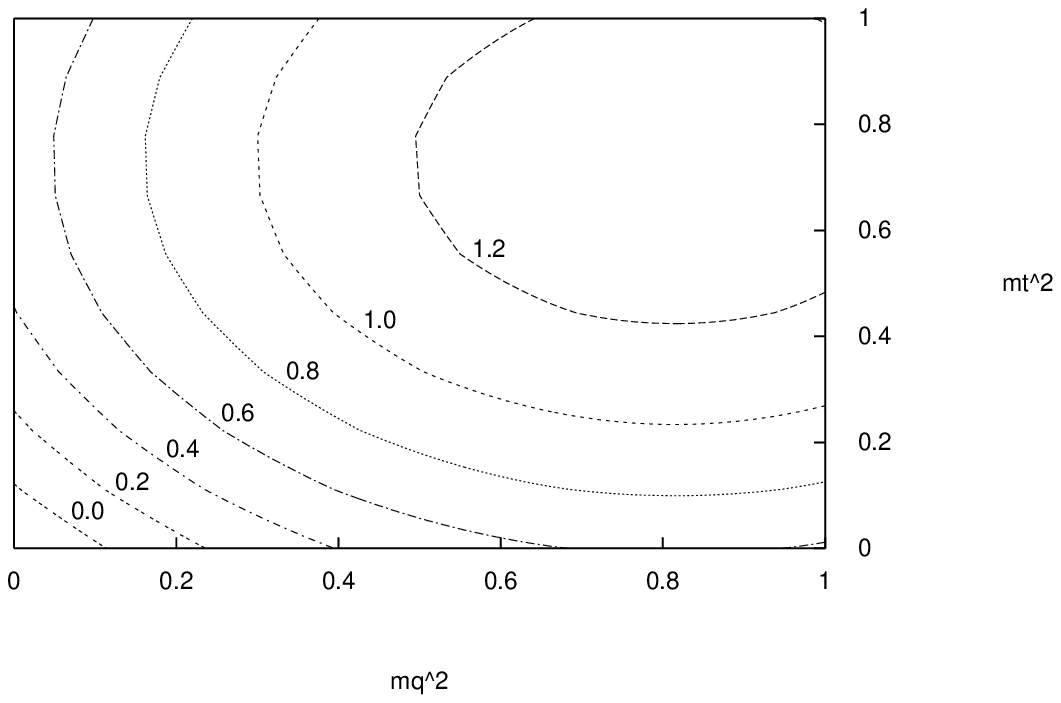}
\caption{One-loop CCB bound for method S given in
Table \protect\ref{method_S_fits}.  $m_{2}^{2}=0.25$ and
each contour is labeled with the corresponding maximum $A$ value.}
\label{loop_S_fig}
\end{figure}

\begin{figure}
\epsfxsize=7.0in
\epsfbox{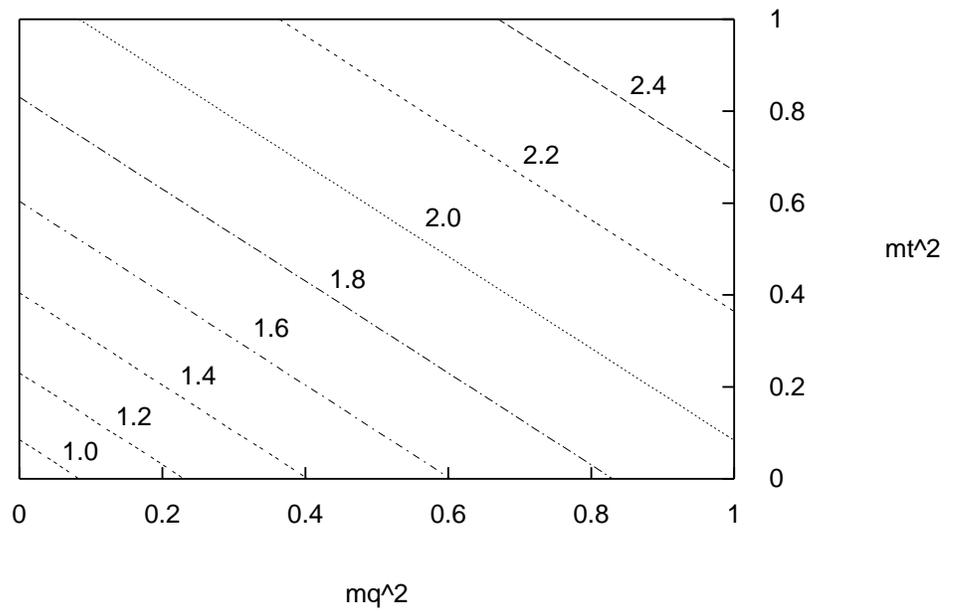}
\caption{Analytical CCB bound for the potential of
Eq. (\protect\ref{GHS_pot}) given in
Eq. (\protect\ref{analytical_bound}).  $m_{2}^{2}=0.25$ and
each contour is labeled with the corresponding $A$ value.}
\label{analytical_fig}
\end{figure}

We present some contour plots of the CCB bounds with
$m_{2}^{2}=0.25$ for comparison.  The contours show the value of
$A_{t}$ on the CCB boundary, i.e., lower values of $A_{t}$ result in a
symmetric vacuum whereas higher values result in a CCB vacuum with
nonzero squark vevs.
Figs. \ref{tree_F_fig} and \ref{loop_F_fig} show the bounds for method
F using the tree-level and one-loop effective potential respectively.
Figs. \ref{tree_S_fig} and \ref{loop_S_fig} show the tree-level and
one-loop method S CCB bounds.  Fig. \ref{analytical_fig} shows the
analytical bound of Eq. (\ref{analytical_bound}) with $m_{2}^{2}=0.25$.

Since there is not the
additional constraint of requiring a Standard Model minimum for the
Higgs field for both the method F CCB bounds and
for the analytical bound,
one may compare these CCB bounds.
The tree-level CCB bound on the $A_{t}$ parameter for method F is
lower than the analytical
bound of Eq. (\ref{analytical_bound}) for the entire range of parameter
values, $(m_{2}^{2},m_{\q}^{2},m_{\tt}^{2})$, shown in
Table \ref{MSSM_range}.  The one-loop correction to
the effective potential raises the value of $A_{t}$ for the CCB
bound over about $70\%$ of the parameter range considered.  However,
even with the one-loop corrections, the CCB bound for the MSSM
potential is more stringent, i.e., gives a lower value for $A_{t}$, than
the bound of Eq. (\ref{analytical_bound}) for $>95\%$ of the parameter range.

The one-loop corrections for the CCB bound calculated using method S
give a lower $A_{t}$ value than the tree-level bound over the entire
range of parameters examined.  For values of the parameters, $(m_{2}^{2},
m_{\q}^{2},m_{\tt}^{2})$, that give a small value for the $A_{t}$ parameter
  CCB bound, the one-loop corrections to the effective potential makes
  the CCB bound significantly stricter.

In conclusion, CCB bounds on the soft-breaking parameters of the Higgs and top
quark/squark sectors of the MSSM provide important constraints for
these parameters.  These constraints may be expressed as a maximum
value of the $A_{t}$ parameter for given values of the remaining
soft-breaking parameters.  The numerical CCB constraints that we
calculated give more stringent CCB bounds than the analytical constraint of
Eq. (\ref{analytical_bound}) for most of the ranges of parameter values
considered.  Because of the large top Yukawa coupling, one-loop
corrections to the effective potential may result in significantly
different CCB bounds than those for the tree-level potential.

\section*{Acknowledgements}
The author wishes to thank Y. Okada for suggesting this topic
and B. Wright for many useful discussions.  This work was partially
supported by the Japan Society for the Promotion of Science.

\section*{Appendix}
\subsection*{Tree-level MSSM Potential}
The potential may be divided into several parts.  A sum over group
and spinor indices, where applicable, is implied.  The supersymmetric
D-terms are
\begin{eqnarray}
{V_{D}} &=& {1\over8}g_{1}^{2}\left(-H_{1}^{\dagger}H_{1}+H_{2}^{\dagger}H_{2}
+ \third|\q|^{2}-{4\over3}|\t|^{2}\right)^{2} \\
&+&{1\over8}g_{2}^{2}\sum_{a=1}^{3}\left(H_{1}^{\dagger}\tau_{a}H_{1}
+ H_{2}^{\dagger}\tau_{a}H_{2} + \q^{*}\tau_{a}\q\right)^{2} \nonumber \\
&+& \half g_{3}^{2}\sum_{a=1}^{8}\left(\q^{*}T^{a}\q
- \t^{*}T^{*a}\t\right)^{2}, \nonumber
\end{eqnarray}
where $g_{1}$, $g_{2}$, and $g_{3}$ are respectively the $U(1)_{Y}$,
$SU(2)$, and $SU(3)$ couplings, $\tau_{a}$ are the Pauli matrices and
$T^{a}$ are the antihermitian generators of $SU(3)$.
Using the relations
\beq
\tau_{ij}^{a}\tau_{kl}^{a} = 2\delta_{il}\delta_{jk}
- \delta_{ij}\delta_{kl}
\eeq
and
\beq
T_{ij}^{a}T_{kl}^{a} = \half\left(\delta_{il}\delta_{jk}
-\third\delta_{ij}\delta_{kl}\right)
\eeq
the $SU(2)$ contribution becomes
\begin{eqnarray}
V_{SU(2)} &=& {1\over 8} g_{2}\Bigl[(H_{1}^{\dagger}H_{1})^{2}
+ (H_{2}^{\dagger}H_{2})^{2} + (\q^{\dagger}\q)^{2} \\
&-& 2\left((H_{1}^{\dagger}H_{1})(H_{2}^{\dagger}H_{2})
+ (H_{1}^{\dagger}H_{1})(\q^{\dagger}\q) + (H_{2}^{\dagger}H_{2})
(\q^{\dagger}\q)\right) \nonumber \\
&+& 4\left(|H_{1}^{\dagger}H_{2}|^{2} + |H_{1}\q|^{2}
+ |H_{2}^{\dagger}\q|^{2}\right)\Bigr] \nonumber
\end{eqnarray}
and the $SU(3)$ one is
\beq
V_{SU(3)} = \half g_{3}^{2}\left[\third(\q^{\dagger}\q)^{2}
+ \third(\t^{\dagger}\t)^{2} - (\q\t)^{\dagger}(\q\t)
+ \third (\q^{\dagger}\q)(\t^{\dagger}\t)\right].
\eeq
The superpotential or F term is
\beq
V_{F} = h_{t}^{2}\left(|\q|^{2}|H_{2}^{0}|^{2} + |\t|^{2}|H_{2}^{0}|^{2}
+ |\q\t|^{2}\right) + h_{t}\mu(\q\t H_{1}^{0*}) + h.c.,
\eeq
with $h.c.$ denoting the Hermitian conjugate and $H_{1}^{0}$ and
$H_{2}^{0}$ are the neutral components of the Higgs
scalar doublets.  The Higgs scalar and fermion doublet components are
\beq
H_{1} = \left( \begin{array}{c}
H_{1}^{0} \\
H_{1}^{-}
\end{array} \right),
\qquad
H_{2} = \left( \begin{array}{c}
H_{2}^{+} \\
H_{2}^{0}
\end{array} \right).
\eeq
The quark-squark-gluino interaction terms are
\beq
V_{q\q\lambda} = \imath\sqrt{2}g_{3}\left(\bt P_{L}\lambda^{(a)}T^{a}\t^{*}
-\t T^{a}\blam^{(a)} P_{R}t + \q^{*}T^{a}\blam P_{L}t
- \bt T^{a}\q P_{R}\lambda^{(a)}\right).
\eeq
$P_{L,R}$ are the projection operators for left- and right-handed
chiral spinors, $\half(1 \pm \gamma^{5})$, $t$ is the four component
spinor field for the top quark, and $\lambda^{(a)}$ are the Majorana
gluino fields.
The quark-squark-Higgsino terms are
\beq
V_{q\q\H} = h_{t}\left(\t\bar{\H}_{2}^{0}P_{L}t
+ \t^{*}\bt P_{R}\H_{2}^{0} - \q\bt P_{L}\H_{2}^{0}
- \q^{*}\bar{\H}_{2}^{0}P_{R}t\right).
\eeq
The Higgsino interaction terms are
\beq
V_{\H\H} = \mu\left(\H_{1}^{0}\H_{2}^{0} - \H_{1}^{-}\H_{2}^{+}\right)
+ h.c.
\eeq
Finally the SUSY soft-breaking terms are
\begin{eqnarray}
V_{soft-breaking} &=& m_{H1}^{2}H_{1}^{\dagger}H_{1}
+ m_{H2}^{2}H_{2}^{\dagger}H_{2} - m_{3}^{2}\epsilon_{ab}
H_{1}^{a}H_{2}^{b} \nonumber \\
&+& m_{\q}^{2}|\q|^{2} + m_{\tt}^{2}|\t|^{2}
+ \half m_{\lambda}\blam^{(a)}\lambda^{(a)}
+ h_{t} A_{t}\q\t H_{2}^{0} + h.c.
\end{eqnarray}
With the addition of the supersymmetric Higgs interactions, the masses for
$H_{1}$ and $H_{2}$ become
\begin{eqnarray}
\label{Higgs_mass_def}
m_{1}^{2} = m_{H1}^{2} + \mu^{2} \\
m_{2}^{2} = m_{H2}^{2} + \mu^{2} \nonumber
\end{eqnarray}
respectively.
\subsection*{One-loop Effective Potential with Zero Squark VEVs}
If the squark vevs are zero the one-loop contributions to the
Higgs effective potential from top squark and quark loops may be
written in an analytical form \cite{analytical form}.  After
requiring that the minimum of the effective potential be at $H_{1} =
v_{1}$ and $H_{2} = v_{2}$ and solving for $m_{1}^{2}$ and $m_{3}^{2}$
we obtain
\begin{eqnarray}
\label{one_loop_m1m3}
m_{3}^{2} &=& (m_{2}^{2} - m_{Z}^{2}\cos 2\beta)\tan\beta \\
&+& {3\over{16\pi^{2}}}\Biggl[(h_{t}^{2}A_{t}(A_{t}\tan\beta + \mu)
{{f(m_{2}^{'2}) - f(m_{1}^{'2})}\over{m_{2}^{'2}-m_{1}^{'2}}} \nonumber \\
&+& h_{t}^{2}\tan\beta(f(m_{1}^{'2})+f(m_{2}^{'2})
-2f(m_{t}^{2}))\Biggr], \nonumber \\
m_{1}^{2} &=& m_{3}^{2}\tan\beta - m_{Z}^{2}\cos 2\beta \nonumber \\
&+& {3\over{16\pi^{2}}}h_{t}^{2}\mu(A_{t}\tan\beta + \mu)
{{f(m_{1}^{'2})-f(m_{2}^{'2})}\over{m_{2}^{'2}-m_{1}^{'2}}}, \nonumber
\end{eqnarray}
with $f(m^{2})\equiv m^{2}(\ln (m^{2}/Q^{2}) - 1)$.
The definition $\tan\beta = v_{2}/v_{1}$ and the tree-level Z boson
mass, $m_{Z}^{2} = (v_{1}^{2} + v_{2}^{2})(g_{1}^{2} + g_{2}^{2})/4$,
were used.  The tree-level relation follows by including only the
first term in the above equations for $m_{1}^{2}$ and $m_{3}^{2}$.

\pagebreak
\begin{table}
\centering
\caption{Grid of parameter values used to calculate the CCB bound for the
Lagrangian of Eq. (\protect\ref{GHS_pot}).}
\label{GHS_range}
\vspace{0.5cm}
\begin{tabular}{cddd}
Parameter & Min. Value & Max. Value & Grid Spacing \\ \hline
$A$ & 0.0 & 3.0 & 0.1 \\
$m_{2}^{2}$ & -0.250 & 0.750 & 0.075 \\
$m_{Q}^{2}$ & 0.0 & 0.750 & 0.075 \\
$m_{T}^{2}$ & 0.0 & 0.750 & 0.075
\end{tabular}
\end{table}

\vspace{1cm}
\begin{table}
\centering
\caption{CCB bounds for the potential of Eq. (\protect\ref{GHS_pot})
  with $h=0.1$ and $h=1.0$.
  The coefficients, $P_{i}$, are defined in Eq.
(\protect\ref{quadratic_surface}).}
\label{GHS_fits}
\vspace{0.5cm}
\begin{tabular}{c|dd}
Parameter & \multicolumn{1}{c}{$h=0.1$} &
\multicolumn{1}{c}{$h=1.0$} \\ \hline
$P_{1}$ & 0.7246 & 2.723 \\
$P_{2}$ & 0.1608 & 3.527 \\
$P_{3}$ & -3.808 & -8.310 \\
$P_{4}$ & 0.1991 & 3.370 \\
$P_{5}$ & -3.830 & -8.228 \\
$P_{6}$ & 0.1985 & 3.452 \\
$P_{7}$ & -3.836 & -8.310 \\
$P_{8}$ & 0.4911 & 0.5877 \\
$\langle d^{2} \rangle$ & 2.$8 \times 10^{-4}$
& 4.$1 \times 10^{-4}$
\end{tabular}
\end{table}

\pagebreak
\vspace*{1cm}
\begin{table}
\centering
\caption{Grid of parameter values used to calculate the tree-level and
  one-loop CCB bounds for the MSSM Lagrangian given in the appendix.}
\label{MSSM_range}
\vspace{0.5cm}
\begin{tabular}{cdddd}
Parameter & Min. Value & Max. Value &
Tree-level Grid Spacing & One-loop Grid Spacing \\ \hline
$A_{t}$ & 0.0 & 3.0 & 0.1 & 0.4 \\
$m_{2}^{2}$ & 2.$5\times 10^{-3}$ & 0.7525 & 0.125 & 0.25 \\
$m_{\q}^{2}$ & 2.$5\times 10^{-3}$ & 0.7525 & 0.125 & 0.25 \\
$m_{\tt}^{2}$ & 2.$5\times 10^{-3}$ & 0.7525 & 0.125 & 0.25
\end{tabular}
\end{table}

\vspace{1cm}
\begin{table}
\centering
\caption{Tree-level and one-loop CCB bounds for the MSSM Lagrangian
using method S.
  The coefficients, $P_{i}$, are defined in Eq.
(\protect\ref{quadratic_surface}).}
\label{method_S_fits}
\vspace{0.5cm}
\begin{tabular}{c|dd}
Parameter & \multicolumn{1}{c}{tree-level} &
\multicolumn{1}{c}{one-loop} \\ \hline
$P_{1}$ & 6.$192 \times 10^{3}$ & 5.$370 \times 10^{1}$ \\
$P_{2}$ & 3.$078 \times 10^{3}$ & 1.$203 \times 10^{1}$ \\
$P_{3}$ & -3.$107 \times 10^{3}$ & -4.$913 \times 10^{0}$ \\
$P_{4}$ &  6.$721 \times 10^{3}$ & 6.$872 \times 10^{1}$ \\
$P_{5}$ &  -1.$162 \times 10^{4}$ & -1.$120 \times 10^{2}$ \\
$P_{6}$ & 6.$857 \times 10^{3}$ & 7.$268 \times 10^{1}$ \\
$P_{7}$ & -1.$175 \times 10^{4}$ & -1.$068 \times 10^{2}$\\
$P_{8}$ & -1.$305 \times 10^{3}$& -1.$233 \times 10^{1}$ \\
$\langle d^{2} \rangle$ & 1.$2 \times 10^{-3}$ &
4.$8 \times 10^{-3}$
\end{tabular}
\end{table}

\pagebreak

\vspace*{1cm}
\begin{table}
\centering
\caption{Tree-level and one-loop CCB bounds for the MSSM Lagrangian
using method F.
  The coefficients, $P_{i}$, are defined in Eq.
(\protect\ref{quadratic_surface}).}
\label{method_F_fits}
\vspace{0.5cm}
\begin{tabular}{c|dd}
Parameter & \multicolumn{1}{c}{tree-level} &
\multicolumn{1}{c}{one-loop} \\ \hline
$P_{1}$ & 4.$431 \times 10^{2}$  & 3.$000 \times 10^{0}$ \\
$P_{2}$ & 9.$397 \times 10^{2}$ & -1.$307 \times 10^{1}$  \\
$P_{3}$ &  -1.$452 \times 10^{3}$ & 5.$948 \times 10^{0}$ \\
$P_{4}$ & 3.$478 \times 10^{2}$ & 7.$767 \times 10^{0}$  \\
$P_{5}$ & -6.$650 \times 10^{2}$ & -9.$086 \times 10^{0}$ \\
$P_{6}$ & 3.$553 \times 10^{2}$ & -7.$102 \times 10^{0}$ \\
$P_{7}$ & -6.$700 \times 10^{2}$ & 3.$374 \times 10^{0}$ \\
$P_{8}$ & -2.$346 \times 10^{2}$ & -2.$833 \times 10^{-1}$ \\
$\langle d^{2} \rangle$ & 1.$6 \times 10^{-3}$
& 9.$4 \times 10^{-3}$
\end{tabular}
\end{table}

\pagebreak

\pagebreak

\vspace*{1.5cm}

\pagebreak

\vspace*{1.5cm}

\pagebreak

\vspace*{1.5cm}

\pagebreak

\vspace*{1.5cm}

\end{document}